\documentclass[twocolumn,showpacs,superscriptaddress,amsmath,amssymb]{revtex4}

\usepackage{graphicx}

\begin{document}

\title{Transport of ultracold Bose gases beyond the Gross-Pitaevskii description}

\author{Thomas Ernst}
\affiliation{Centre for Theoretical Chemistry and Physics and Institute of
	Natural Sciences, Massey University, Private Bag 102 904, North Shore,
	Auckland 0745, New Zealand}
\author{Tobias Paul}
\affiliation{Institut f\"ur Theoretische Physik, Universit\"at Heidelberg,
Philosophenweg 19, 69120 Heidelberg, Germany}
\author{Peter Schlagheck}
\affiliation{Institut f\"ur Theoretische Physik, Universit\"at Regensburg,
  93040 Regensburg, Germany}
\affiliation{Mathematical Physics, Lund Institute of Technology, PO Box 118,
22100 Lund, Sweden}

\date{\today}

\begin{abstract}

We explore atom-laser-like transport processes of ultracold Bose-condensed
atomic vapors in mesoscopic waveguide structures beyond the Gross-Pitaevskii
mean-field theory. 
Based on a microscopic description of the transport process in the presence of
a coherent source which models the outcoupling from a reservoir of perfectly
Bose-Einstein condensed atoms, we derive a system of coupled quantum evolution
equations that describe the dynamics of a dilute condensed Bose gas in the
framework of the Hartree-Fock-Bogoliubov approximation.
We apply this method to study the transport of dilute Bose gases through an
atomic quantum dot and through waveguides with disorder. Our numerical
simulations reveal that the onset of an explictly time-dependent flow
corresponds to the appearance of strong depletion of the condensate on the
microscopic level and leads to a loss of global phase coherence.

\end{abstract}


\maketitle

\section{\label{intro}Introduction}

The rapid progress in the experimental techniques for trapping and
manipulating ultracold Bose gases on microscopic scales has opened the
possibility for investigating mesoscopic transport properties with interacting
bosonic matter waves.
A key achievement in this context is the development of simple or more complex
waveguide geometries for cold atoms with optical techniques (as, e.g., in
Ref.~\cite{DumO02PRL}) or with atom chips \cite{FolO00PRL,HaeO01N,ForZim07RMP},
which also allow one to impose scattering and disorder potentials on
microscopic scales (e.g.\ \cite{BilO08N,RoaO08N}) and which also permit to
detect individual atoms with rather good accuracy \cite{CamO06PRA}.
Coherent transport processes in such waveguides, where the atoms of the Bose
gas are freely propagating along the guide with some finite momentum, can be 
studied, on the one hand, through bosonic wave packets that are created upon 
sudden release of a Bose-Einstein condensate from a trap, which is then
accelerated in the presence of a finite potential gradient along the guide
\cite{OttO03PRL,LyeO07PRA}.
On the other hand, the principle of an atom laser
\cite{MewO97PRL,BloHaeEss99PRL} can be used for this purpose, as was recently
demonstrated in Refs.~\cite{GueO06PRL,BilO07AP,RioO08PRA,CouO08EPL}.
In these experiments, a coherent matter-wave beam was injected from a
Bose-Einstein condensate in a trap into an optical waveguide.
This was achieved by means of an radio-frequency (rf) induced transition from
a magnetically trapped ($m_F = -1$) to an untrapped ($m_F = 0$) hyperfine
state of $^{87}$Rb in Ref.~\cite{GueO06PRL,BilO07AP}, and by means of a
careful ramping of a magnetic-field gradient in an optical trap in
Ref.~\cite{CouO08EPL}.
In this way it becomes possible to study bosonic scattering processes with
cold atoms at well-defined incident energy, in close analogy to scattering of
laser beams and to electronic transport in mesoscopic solid-state systems
\cite{FerGoo}.

From the theoretical side, a number of investigations on the quasicontinuous
transport of ultracold Bose gases and Bose-Einstein condensates have been
undertaken during the past decade.
This started with the attempt to define an atomic analog of Landauer's
quantization of conductance \cite{ThyWesPre99PRL} and was continued by first
investigations of nonlinear resonant transport and interaction blockade in
quantum-dot-like scattering potentials \cite{CarLar00PRL,Car01PRA}.
Propagation and transmission studies were then undertaken on the basis of the
stationary Gross-Pitaevskii equation
\cite{LebPav01PRA,LebPavSin03PRA,RapWitKor06PRA,RapKor08PRA} as well as
through a time-dependent integration approach in which the injection of
bosonic matter waves was accounted for by means of a coherent source term
\cite{PauRicSch05PRL,PauO05PRA,PauO07PRA}.

With few exceptions (e.g., Ref.~\cite{Car01PRA}), the above studies were
mainly based on an effective one-dimensional Gross-Pitaevskii equation.
This implicitly assumes the realization of a one-dimensional mean-field regime
\cite{MenStr02PRA} where the transverse confinement of the waveguide is strong
enough to inhibit the population of transversally excited modes, but not so
strong as to come close to the Tonk-Girardeau regime of impenetrable atoms
\cite{Gir60JMP,Ols98PRL}.
Furthermore, the phase coherence length for a freely propagating atomic beam,
which is generally finite for one-dimensional quasi-condensates
\cite{PetShlWal00PRL}, is assumed to be considerably larger than the effective
longitudinal extent of the waveguide (which may be limited due to a finite
focal region in the case of an elongated dipole guide \cite{GueO06PRL}, or due
to atom detection \cite{CamO06PRA}).
Using the one-dimensional Gross-Pitaevskii equation under these conditions,
the presence of atom-atom interaction in the bosonic beam mainly manifests, as
in nonlinear optics \cite{Boy}, in form finite nonlinearity effects in the
wave  scattering process, such as nonlinearity-induced shifts of resonant
transmission peaks \cite{CarLar00PRL,PauRicSch05PRL,RapWitKor06PRA}.

The main focus of this work is to determine the quantitative amount of
\emph{depletion} that is generated during the transport process of the
condensate, which is a rather relevant information from the experimental point
of view.
Indeed, too much depletion would eventually lead to a complete destruction of
the phase coherence of the beam, which means that wave interference phenomena
in the transport process, which would result from the coherent Gross-Pitaevskii
equation, may, in this case, not be observable in practice.
Inspired by previous studies \cite{ProBur96JRNIST,KoeBur02PRA,HutO00JPB},
we use a quantum kinetic approach for this purpose, which is equivalent to the
Hartree-Fock-Bogoliubov approximation \cite{Gri96PRB}.
In practice, the condensate wavefunction is, in this approach, propagated
together with two-component functions that describe the coherence as well as
the one-body density matrix associated with non-condensed atoms.
This ansatz generally involves a renormalization of the microscopic
interaction strength, which, however, does not represent a conceptual
problem in the quasi-one-dimensional confinement geometries that we are
considering here.
Moreover, it explicitly breaks the gauge symmetry of the Bose gas and
introduces a gap in the excitation spectrum \cite{Gri96PRB,HutO00JPB}.
We justify this due to the fact that the matter-wave beam is, in the guided
atom-laser scenario, connected to an idealized reservoir that contains a
macroscopically large Bose-Einstein condensate, for which spontaneous breaking
of gauge symmetry can be well assumed \cite{LieSeiYng07RMathP}.

In order to provide a solid theoretical foundation of our approach, we begin,
in Section \ref{sec:eq}, with a step-by-step derivation of the quantum
kinetic equations that are used to model the propagation process of the
condensate in the presence of depletion.
For the sake of definiteness, we focus here on an atom laser process that is
based on the rf-induced outcoupling of atoms from a Bose-Einstein condensate
in a magnetic trap \cite{GueO06PRL,BilO07AP}.
Our starting point is the microscopic many-body description of the
two-component system that defines this atom laser, from which we derive,
under the assumption of a weak effective atom-atom interaction in the
waveguide, nonlinear kinetic equations for the condensate wavefunction
and for two-component functions that account, in lowest order, for quantum
depletion.
In Sections \ref{sec:db} and \ref{sec:dis}, respectively, we apply this
approach to transport processes of Bose-Einstein condensates through double
barrier potentials and disorder potentials that were also studied in
Refs.~\cite{PauRicSch05PRL,PauO05PRA} on the basis of the Gross-Pitaevskii
equation.
Our main finding, which is summarized in the conclusion in Section
\ref{sec:concl}, is that the overall coherence of the atomic beam remains
fairly well preserved over reasonably long time scales in the case of a
quasi-stationary flow of the condensate, while strong depletion arises in the
case of permanently time-dependent scattering in the waveguide.

\section{Kinetic equations for the scattering system}

\label{sec:eq}

We consider a gas of ultracold bosonic atoms that are confined in a
large trap and injected from there into a mesoscopic waveguide structure.
In practice, such a ``guided atom laser'' can be realized by combining a
magnetic trapping potential with an optical waveguide which is, e.g., created
by an elongated dipole trap.
A radio-frequency (rf) field is then used to flip the spins of the atoms and
induce coherent transitions from a low-field seeker state $|r\rangle$ (such as
the $|F,m_F\rangle = |1, -1 \rangle$ state of $^{87}$Rb) to a hyperfine state
$|g\rangle$ that is, in lowest order, insensitive to the magnetic field (the
$|F,m_F\rangle = |0, 0 \rangle$ state of $^{87}$Rb) but still experiences the
optical potential.
As demonstrated in Refs.~\cite{GueO06PRL,BilO07AP} where this setup was
experimentally realized, the quadratic Zeeman effect can be exploited to
compensate the quadratic variation of the laser intensity along the dipole
trap and to thereby create a rather homogeneous waveguide potential.

In the formalism of second quantization, the quantum dynamics of this
two-component system is described by the coupled equations
\begin{eqnarray}
  i \hbar \frac{\partial}{\partial t} \hat{\Psi}_g(\mathbf{r},t) & = &
  \left( - \frac{\hbar^2}{2m} \Delta + V_g(\mathbf{r}) \right) 
  \hat{\Psi}_g(\mathbf{r},t) \nonumber \\
  & & + U_g \hat{\Psi}_g^\dagger(\mathbf{r},t) \hat{\Psi}_g(\mathbf{r},t) 
  \hat{\Psi}_g(\mathbf{r},t) \nonumber \\
  & & + U_{r g} \hat{\Psi}_r^\dagger(\mathbf{r},t) \hat{\Psi}_r(\mathbf{r},t)
  \hat{\Psi}_g(\mathbf{r},t) \nonumber \\
  & & + K(t) \hat{\Psi}_r(\mathbf{r},t) \label{eq:fullguide} \\
  i \hbar \frac{\partial}{\partial t} \hat{\Psi}_r(\mathbf{r},t) & = &
  \left( - \frac{\hbar^2}{2m} \Delta + V_r(\mathbf{r}) \right) 
  \hat{\Psi}_r(\mathbf{r},t) \nonumber \\
  & & + U_r \hat{\Psi}_r^\dagger(\mathbf{r},t) \hat{\Psi}_r(\mathbf{r},t) 
  \hat{\Psi}_r(\mathbf{r},t) \nonumber \\
  & & + U_{r g} \hat{\Psi}_g^\dagger(\mathbf{r},t) \hat{\Psi}_g(\mathbf{r},t)
  \hat{\Psi}_r(\mathbf{r},t) \nonumber \\
  & & + K^*(t) \hat{\Psi}_g(\mathbf{r},t) \label{eq:fullres} 
\end{eqnarray}
for the bosonic field operators $\hat{\Psi}_g(\mathbf{r},t)$ and
$\hat{\Psi}_r(\mathbf{r},t)$ that annihilate atoms in the ``waveguide state''
$|g\rangle$ and the ``reservoir state'' $|r\rangle$, respectively.
$V_g(\mathbf{r})$ denotes the waveguide potential while $V_r(\mathbf{r})$
corresponds to the magnetic trap (perturbed by the presence of the
optical guide) that confines the reservoir atoms in the state $|r\rangle$.
The short-range interaction between the atoms in the involed hyperfine states
is represented by contact potentials of the form 
$U(\mathbf{r}_1-\mathbf{r}_2) \propto \delta(\mathbf{r}_1-\mathbf{r}_2)$ 
with the prefactors $U_g$, $U_r$, and $U_{r g}$ for $|g\rangle$-$|g\rangle$, 
$|r\rangle$-$|r\rangle$, and $|g\rangle$-$|r\rangle$ interaction processes,
respectively.
Transitions between waveguide and reservoir states are induced by the coupling
amplitude $K(t)$, which is adiabatically raised from zero to a finite value
and which would correspond to the field strength of the rf radiation.

It is reasonable from the experimental point of view to consider an initial
many-body state in the trap that consists of a nearly perfect Bose-Einstein
condensate containing a rather large number of atoms.
Provided the coupling amplitude $K(t)$ is not too strong, we can safely assume
that the condensate is not appreciably affected by the outcoupling process
on finite time scales.
We therefore make the ansatz
\begin{equation}
  \hat{\Psi}_r(\mathbf{r},t) = \langle \hat{\Psi}_r(\mathbf{r},t) \rangle
  = \Psi_r(\mathbf{r}) \exp(-i \mu t / \hbar)
\end{equation}
where $\Psi_r(\mathbf{r})$ is the macroscopically populated condensate
wavefunction that satisfies the Gross-Pitaevskii equation
\begin{equation}
  \left( - \frac{\hbar^2}{2m} \Delta + V_r(\mathbf{r}) + U_r
    |\Psi_r(\mathbf{r})|^2 \right) \Psi_r(\mathbf{r}) = \mu \Psi_r(\mathbf{r})
\end{equation}
with the normalization
\begin{equation}
	\int d^3r |\Psi_r(\mathbf{r})|^2 = \mathcal{N}_r \gg 1,
\end{equation}
and $\mu$ denotes the chemical potential of the condensate.
This ansatz explicitly involves a spontaneous breaking of the condensate's
gauge symmetry, which should be valid in the thermodynamic limit of a very
large number $\mathcal{N}_r$ of atoms in the trap \cite{LieSeiYng07RMathP}.
Inserting this expression for $\hat{\Psi}_r(\mathbf{r},t)$ into
Eq.~(\ref{eq:fullguide}) yields then
\begin{eqnarray}
  i \hbar \frac{\partial}{\partial t} \hat{\Psi}_g(\mathbf{r},t) & = &
  \left( - \frac{\hbar^2}{2m} \Delta + V_g(\mathbf{r}) + U_{r g}
    |\Psi_r(\mathbf{r})|^2 \right) \hat{\Psi}_g(\mathbf{r},t) \nonumber \\
  & & + U_g \hat{\Psi}_g^\dagger(\mathbf{r},t) \hat{\Psi}_g(\mathbf{r},t) 
  \hat{\Psi}_g(\mathbf{r},t) \nonumber \\
  & & + K(t) \Psi_r(\mathbf{r}) \exp(-i \mu t / \hbar) \label{eq:guide}
\end{eqnarray}
as equation for the field operator of the waveguide state, which contains
a coherent source term with the amplitude $K(t) \Psi_r(\mathbf{r})$.

The external potential for the atoms in the state $|g\rangle$ is written as
\begin{equation}
  V_g(\mathbf{r}) = \frac{1}{2} m \omega_\perp^2 r_\perp^2 + V(x)
\end{equation}
where $x$ and $\mathbf{r}_\perp \equiv (y,z)$ denote the spatial coordinates
along and perpendicular to the waveguide, respectively, and $\omega_\perp$
is the transverse confinement frequency.
In addition to the waveguide, we consider the presence of a longitudinal
scattering potential $V(x)$ of finite spatial range, which could, e.g., be a
sequence of barriers or speckle disorder induced by other laser fields.
Denoting the transverse eigenstates within the waveguide by 
$\chi_n(\mathbf{r}_\perp)$, we can make the decomposition
$\hat{\Psi}_g(\mathbf{r},t) = \sum_n \chi_n(\mathbf{r}_\perp)
\hat{\psi}_n(x,t)$ for the field operator of the $|g\rangle$-atoms, where
$\hat{\psi}_n(x,t)$ annihilates a particle at position $x$ in the $n$th
transverse eigenmode.
This decomposition can be generalized for waveguides with confinement
frequencies $\omega_\perp$ that are slowly varying with the longitudinal
coordinate $x$;
in that case, the eigenstates $\chi_n(\mathbf{r}_\perp)$ would vary with $x$ as
well, through their parametric dependence on $\omega_\perp$ \cite{JaeSte02PRA}.

\begin{figure}
\includegraphics[width=1.0\columnwidth]{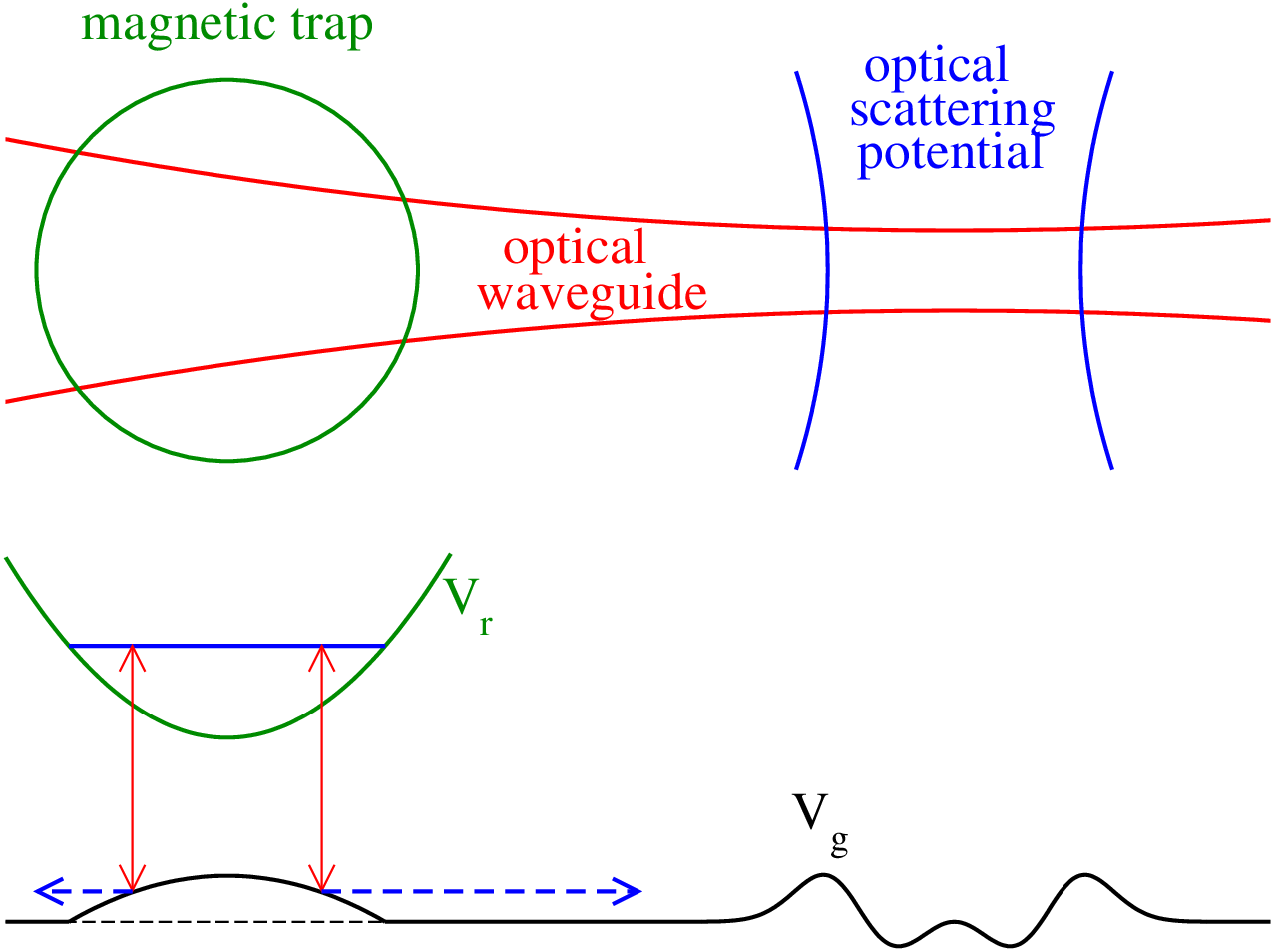}
\caption{\label{fig:source} (Color online)
Possible waveguide geometry for a guided atom laser experiment with
Bose-Einstein condensates in the presence of an optical scattering potential.
The lower panel shows the effective potentials along the longitudinal
coordinate $x$ for the reservoir ($V_r$; green parabola) and the waveguide
atoms ($V_g$; black horizontal curve).
A radiofrequency field with suitable frequency $\omega_{\rm rf}$ (indicated 
by the vertical red line) can be used to resonantly couple atoms from the 
trapped condensate (whose chemical potential is denoted by the horizontal
blue line) into the waveguide.
}
\end{figure}

In general, the coherent source in Eq.~(\ref{eq:guide}) populates a finite
linear combination of transverse eigenmodes in the waveguide, which would be
coupled to each other through the interaction term.
Clean bosonic scattering experiments, however, would ideally require a
well-defined longitudinal kinetic energy of the atoms, which means that the
population ought to be restricted to one single transverse mode.
Technically, this task could possibly be accomplished by imposing a smooth
barrier potential along the waveguide with a maximum height $V_{\rm max}$ that
satisfies $\mu - 2 \hbar \omega_\perp < V_{\rm max} < \mu - \hbar
\omega_\perp$.
In this case, atoms in the transverse ground mode (with the offset energy 
$\hbar \omega_\perp$) would have just enough kinetic energy to pass that
barrier, while components associated with excited transverse modes (with
offset energies $\geq 2 \hbar \omega_\perp$) would be reflected.

In the following, we assume that the population of the waveguide by the source
can indeed be restricted to the transverse ground mode
$\chi_0(\mathbf{r}_\perp) \propto \exp(- m \omega_\perp r_\perp^2 / 2 \hbar)$.
We then make the simplifying ansatz
\begin{equation}
  \hat{\Psi}_g(\mathbf{r},t) = \chi_0(\mathbf{r}_\perp) \hat{\psi}(x,t)
\end{equation}
which restricts the consideration to this ground mode and neglects the effect
of (possibly virtual) transitions to excited transverse modes due to the
interaction.
$\hat{\psi}(x,t)$ then satisfies the equation
\begin{eqnarray}
  i \hbar \frac{\partial}{\partial t} \hat{\psi}(x,t) & = &
  H_0^{(x)} \hat{\psi}(x,t) + g \hat{\psi}^\dagger(x,t) \hat{\psi}(x,t)
  \hat{\psi}(x,t) \nonumber \\
  & & + S(x,t) \exp(-i \mu t / \hbar) \label{eq:1Dguide}
\end{eqnarray}
with the one-dimensional single-particle Hamiltonian
\begin{equation}
  H_0^{(x)} = - \frac{\hbar^2}{2m} \frac{\partial^2}{\partial x^2} + V(x) +
  \hbar \omega_\perp
\end{equation}
and the effective one-dimensional interaction strength
\begin{equation}
  g = \frac{m \omega_\perp}{2 \pi \hbar} U_g = 2 \hbar \omega_\perp a_s
  \label{eq:g}
\end{equation}
where $a_s$ is the $s$-wave scattering length for atoms in the state
$|g\rangle$.
The source amplitude in this one-dimensional equation is given by
\begin{equation}
  S(x,t) = \int d^2r_\perp \chi_0^*(\mathbf{r}_\perp) \Psi_r(\mathbf{r}) K(t).
\end{equation}
Assuming that the condensate wavefunction $\Psi_r(\mathbf{r})$ has no overlap
with the region in which the scattering potential $V(x)$ is defined, we make
in the following the idealized ansatz of a strongly localized source, namely
\begin{equation}
  S(x,t) = S_0(t) \delta(x-x_0) \label{eq:pointsource}
\end{equation}
where $x_0$ corresponds to the position around which the trap is centered.

Our aim is to compute the time evolution of the condensate wavefunction in the
waveguide as well as the amount of depletion that is generated during this
transport process.
To this end, we follow the lines of the treatment in 
Refs.~\cite{ProBur96JRNIST,KoeBur02PRA} and make the Bogoliubov ansatz for
the field operator $\hat{\psi}(x,t)$ in the transverse ground mode of the
waveguide, which is decomposed into its order parameter 
$\langle \hat{\psi}(x,t) \rangle$ and a quantum fluctuation operator 
$\delta \hat{\psi}(x,t)$ according to
\begin{equation}
  \hat{\psi}(x,t) = \langle \hat{\psi}(x,t) \rangle + \delta \hat{\psi}(x,t).
\end{equation}
Using Eq.~(\ref{eq:1Dguide}), it is then straightforward to derive an infinite
set of coupled kinetic equations for the condensate wavefunction 
$\langle \hat{\psi}(x,t) \rangle$ and the expectation values of products of
the operators  $\delta \hat{\psi}(x,t)$ and their hermitean conjugates ---
i.e., for the $n$-point functions
$\langle \delta \hat{\psi}(x_1,t) \delta \hat{\psi}(x_2,t) \rangle$,
$\langle \delta \hat{\psi}^\dagger(x_1,t) \delta \hat{\psi}(x_2,t) \rangle$,
$\langle \delta \hat{\psi}(x_1,t) \delta \hat{\psi}(x_2,t) 
\delta \hat{\psi}(x_3,t) \rangle$, etc.
The time evolution of $\langle \hat{\psi}(x,t) \rangle$, for instance, is
described by the equation
\begin{eqnarray}
  i \hbar \frac{\partial}{\partial t} \langle \hat{\psi}(x,t) \rangle 
  & = & H_0^{(x)} \langle \hat{\psi}(x,t) \rangle + g 
  |\langle \hat{\psi}(x,t) \rangle|^2 \langle \hat{\psi}(x,t) \rangle 
  \nonumber \\
  & & + 2 g \langle \delta \hat{\psi}^\dagger(x,t) \delta \hat{\psi}(x,t) 
  \rangle \langle \hat{\psi}(x,t) \rangle \nonumber \\
  & & + g \langle \delta \hat{\psi}(x,t) \delta \hat{\psi}(x,t) \rangle
  \langle \hat{\psi}^\dagger(x,t) \rangle \nonumber \\
  & & + g \langle \delta \hat{\psi}^\dagger(x,t) \delta \hat{\psi}(x,t) 
  \delta \hat{\psi}(x,t) \rangle \nonumber \\
  & & + S(x,t) \exp(-i \mu t / \hbar), \label{eq:psi0}
\end{eqnarray}
which contains couplings to $n$-point functions up to $n=3$.
The equations for the higher-order cumulants, e.g.\ for 
$\langle \delta \hat{\psi}(x_1,t) \delta \hat{\psi}(x_2,t) \rangle$,
are obtained by the time derivative of $\delta \hat{\psi}(x,t) \equiv
\hat{\psi}(x,t) - \langle \hat{\psi}(x,t) \rangle$ through the combination of
Eqs.~(\ref{eq:1Dguide}) and (\ref{eq:psi0}).

In order to define a meaningful truncation scheme for this infinite hierarchy
of equations, we make the standard mean-field assumption that we have a
rather small interaction strength $g$, corresponding to a small $s$-wave
scattering length $a_s$ of the atoms in the waveguide, and a comparatively
large source amplitude $S_0$, giving rise to a large longitudinal density 
$n_0 = (m / 2 \mu) |S_0 / \hbar|^2$ of atoms that are injected by the source.
The mean-field regime could then be expressed by the inequalities
$n_0 \lambda \gg 1$ and $a_s / \lambda \ll 1$ where $\lambda$ characterizes
the relevant length scales of the waveguide system [such as the transverse
oscillator length $a_\perp = \sqrt{\hbar/m \omega_\perp}$, the de Broglie
wavelength of the incident matter-wave beam, or the length scales introduced
by the scattering potential $V(x)$]. 
Formally, it would be defined by the simultaneous limits $n_0 \to \infty$ and
$a_s \to 0$ with $a_s n_0$ remaining constant \cite{LieSeiYng00PRA}.

This formal mean-field limit allows us now to classify the individual terms in
Eq.~(\ref{eq:psi0}) and in the equations for the higher-order $n$-point
functions according to different powers in the large parameter 
$N \equiv n_0 \lambda$ characterizing the number of atoms within the length
scale $\lambda$.
As we have $|S_0| = \hbar \sqrt{2 \mu n_0 / m} \propto \sqrt{N}$, we can
obviously infer from Eq.~(\ref{eq:psi0}) that the condensate wavefunction 
$\langle \hat{\psi}(x,t) \rangle$ scales as $\mathcal{O}(N^{1/2})$ for finite
evolution times.
The equation for $\langle \delta \hat{\psi}(x_1,t) \delta \hat{\psi}(x_2,t)
\rangle$ [see Eq.~(\ref{eq:phi}) below] is not directly connected to the
coherent source, but contains the inhomogeneous term 
$g \delta(x_1 - x_2) \langle \hat{\psi}(x_1,t) \rangle 
\langle \hat{\psi}(x_2,t) \rangle$ through which this two-point function
becomes populated in the course of time evolution.
Taking into account that $g \propto 1/N$ in this formal mean-field limit
and using $\langle \hat{\psi}(x,t) \rangle \sim \mathcal{O}(N^{1/2})$, we
then obtain $\langle \delta \hat{\psi}(x_1,t) \delta \hat{\psi}(x_2,t)
\rangle \sim \mathcal{O}(N^{0})$.
Similarly one can show $\langle \delta \hat{\psi}^\dagger(x_1,t) 
\delta \hat{\psi}(x_2,t) \rangle \sim \mathcal{O}(N^{0})$ as well, while
three-point functions such as $\langle \delta \hat{\psi}(x_1,t) 
\delta \hat{\psi}(x_2,t) \delta \hat{\psi}(x_3,t) \rangle$ would scale as
$\mathcal{O}(N^{-1/2})$.

Following the notation introduced by K\"ohler and Burnett \cite{KoeBur02PRA},
we define the condensate wavefunction
\begin{equation}
  \psi(x,t) \equiv \langle \hat{\psi}(x,t) \rangle,
\end{equation}
for atoms in the transverse ground mode of the waveguide, the pair function
\begin{equation}
  \Phi(x_1,x_2,t) \equiv
  \langle \delta \hat{\psi}(x_2,t) \delta \hat{\psi}(x_1,t) \rangle,
\end{equation}
and the one-body density matrix of noncondensed atoms
\begin{equation}
  \Gamma(x_1,x_2,t) \equiv
  \langle \delta \hat{\psi}^\dagger(x_2,t) \delta \hat{\psi}(x_1,t) \rangle.
\end{equation}
By taking into account all terms that scale at least as $\mathcal{O}(N^{-1})$,
we obtain a closed system of coupled equations for $\psi(x,t)$,
$\Phi(x_1,x_2,t)$, and $\Gamma(x_1,x_2,t)$, namely
\begin{widetext}
\begin{eqnarray}
  i\hbar\frac{\partial}{\partial t}\psi(x,t) & = &
  H_0^{(x)}\psi(x,t)+g\left|\psi(x,t)\right|^{2}\psi(x,t) 
  + S(x,t) \exp(-i \mu t / \hbar) \hspace{3cm} 
	\left[\mathcal{O}(N^{1/2})\right] \nonumber \\
  & & +2g\psi(x,t)\Gamma(x,x,t)+g\psi^*(x,t)\Phi(x,y,t)
  \hspace{5cm} \left[\mathcal{O}(N^{-1/2})\right] \label{eq:psi} \\
  {i\hbar\frac{\partial}{\partial t}\Phi(x_1,x_2,t)} & = & 
  {\left[H_0^{(x_1)}+2g \left|\psi(x_1,t)\right|^2 + H_0^{(x_2)}+
      2g \left|\psi(x_2,t)\right|^2 \right]\Phi(x_1,x_2,t)} \nonumber \\
  & & {+g\delta(x_1-x_2)\psi(x_1,t)\psi(x_2,t)}\nonumber \\
  & & {+g\psi^2(x_2,t)\Gamma(x_1,x_2,t)+g\psi^2(x_1,t)\Gamma(x_2,x_1,t)
    \hspace{4cm} \left[\mathcal{O}(N^0)\right]}\nonumber \\
  & & {+ g\delta(x_1-x_2)\Phi(x_1,x_2,t) + 
    2g \left[\Gamma(x_1,x_1,t)+\Gamma(x_2,x_2,t)\right]\Phi(x_1,x_2,t)} 
  \nonumber \\ 
  & & {+g\Gamma(x_2,x_1,t)\Phi(x_1,x_1,t)+g\Gamma(x_1,x_2,t)\Phi(x_2,x_2,t)
    \hspace{3.5cm} \left[\mathcal{O}(N^{-1})\right]} \label{eq:phi} \\
  {i\hbar\frac{\partial}{\partial t}\Gamma(x_1,x_2,t)} & = & 
  {\left[H_0^{(x_1)}+2g \left|\psi(x_1,t)\right|^2 -H_0^{(x_2)} - 
      2g \left|\psi(x_2,t)\right|^2\right]\Gamma(x_1,x_2,t)} \nonumber \\
  & & {+g\psi^2(x_1,t)\Phi^{\star}(x_1,x_2,t) - 
    g\left(\psi^{\star}(x_2,t)\right)^2\Phi(x_1,x_2,t) 
    \hspace{3.5cm} \left[\mathcal{O}(N^0)\right]} \nonumber \\
  & & {+2g\left[\Gamma(x_1,x_1,t)-\Gamma(x_2,x_2,t)\right]
    \Gamma(x_1,x_2,t)} \nonumber \\
  & & {+g\Phi^{\star}(x_1,x_2,t)\Phi(x_1,x_1,t) - 
    g\Phi(x_1,x_2,t)\Phi^{\star}(x_2,x_2,t) \hspace{3cm} 
	\left[\mathcal{O}(N^{-1})\right]}
  \label{eq:gamma}
\end{eqnarray}
\end{widetext}
where consistently all couplings to higher-order $n$-point functions
[which would scale as $\mathcal{O}(N^{-3/2})$ in Eq.~(\ref{eq:psi}) and
as $\mathcal{O}(N^{-2})$ in Eqs.~(\ref{eq:phi},\ref{eq:gamma})] are
omitted.

This set of kinetic equations is equivalent to the second-order
Hartree-Fock-Bogoliubov approximation in modal form \cite{Gri96PRB}.
It is norm-conserving, i.e., for $x \neq x_0$ the continuity equation
\begin{equation}
  \frac{\partial}{\partial t} n(x,t) + \frac{\partial}{\partial x} j(x,t) = 0
\end{equation}
is satisfied with
\begin{eqnarray}
  n(x,t) & \equiv & \langle \hat{\psi}^\dagger(x,t) \hat{\psi}(x,t) \rangle
  \nonumber \\
  & = & |\psi(x,t)|^2 + \Gamma(x,x,t)
\end{eqnarray}
the longitudinal particle density and
\begin{eqnarray}
  j(x,t) & \equiv & \frac{\hbar}{2 m i} \left. \left( 
      \frac{\partial}{\partial x_1} - \frac{\partial}{\partial x_2} \right) 
    \langle \hat{\psi}^\dagger(x_2,t) \hat{\psi}(x_1,t) \rangle 
  \right|_{x_1 = x_2 = x} \nonumber \\
  & = & \frac{\hbar}{2 m i} \left[ \psi^*(x,t) \frac{\partial}{\partial x}
    \psi(x,t) - \psi(x,t) \frac{\partial}{\partial x} \psi^*(x,t) \right.
  \nonumber \\
  & & + \left. \left. \left( \frac{\partial}{\partial x_1} - 
      \frac{\partial}{\partial x_2} \right) \Gamma(x_1, x_2) 
  \right|_{x_1 = x_2 = x} \right] \label{eq:current}
\end{eqnarray}
the atomic current along the waveguide.
As for the density $n(x,t)$, the current $j(x,t)$ consists of a coherent
component associated with the condensate and an incoherent component that
represents the noncondensed fraction, given by the second and third line of
Eq.~(\ref{eq:current}), respectively.

The transmission of a scattering state that is associated with a given source
amplitude $S_0$ is obtained by dividing $j(x,t)$ through the incident current
$j_i$. 
The latter is defined as the current that would be obtained at the same source
amplitude $S_0$ if both the scattering potential $V(x)$ and the interaction
strength $g$ were zero.
We straightforwardly obtain for that case the asymptotic solution
\begin{equation}
  \psi(x,t) = - \frac{i S_0}{\hbar} \sqrt{\frac{m}{2 \mu}}
  \exp\left[ \frac{i}{\hbar} \left(\sqrt{2 m \mu} |x - x_0| - \mu t \right) 
  \right]
\end{equation}
which yields
\begin{equation}
  j_i = \left|\frac{S_0}{\hbar}\right|^2 \sqrt{\frac{m}{2 \mu}} \, .
\end{equation}
We point out that the interaction strength $g$ is assumed to vanish at 
$x=x_0$, which avoids effects of nonlinear backaction with the source.
In between the source and the scattering potential, $g$ is adiabatically
increased with $x$ up to a given maximal value, which remains then constant
along the support of $V(x)$.

For the numerical treatment of the kinetic equations
(\ref{eq:psi}-\ref{eq:gamma}), we employ a finite-difference representation of
the one- and two-point functions $\psi(x,t)$, $\Phi(x_1,x_2,t)$, and
$\Gamma(x_1,x_2,t)$ \cite{PauO07PRA}.
The numerical integration of the resulting vectorial differential equations
of the form $i \dot{\mathbf{y}} = \mathcal{H} \mathbf{y}$ is then performed by
means of the Crank-Nicolson method \cite{NumericalRecipes1986}, which
utilizes matrix-vector products as well as the solution of linear
matrix-vector equations involving the effective Hamiltonian matrix
$\mathcal{H}$.
The split-operator technique is applied in the equations for $\Phi$ and
$\Gamma$ in order to do separate propagation steps in the $x_1$ and $x_2$
coordinates.
To account for the nonlinearities, i.e.\ the dependence of $\mathcal{H}$ on
$\mathbf{y}$, each integration step is ``corrected'' by repeating it with
$\mathcal{H}$ being evaluated at $[\mathbf{y}(t) + 
\mathbf{y}^{(1)}(t+\delta t)]/2$ where $\mathbf{y}^{(1)}(t+\delta t)$ denotes
the first-order estimate for $\mathbf{y}(t+\delta t)$
\cite{CerO98PLA,PauO07PRA}. 

In this grid representation, the Dirac delta functions appearing in the source
term (\ref{eq:pointsource}) and in the interaction [cf.\ the second line of
Eq.~(\ref{eq:phi})] are naturally replaced by Kronecker deltas with the
appropriate prefactors.
We verified that the regularization of the interaction potential by a narrow
Gaussian function does not significantly change our numerical results.
The amplitude of the source is adiabatically ramped from zero to $S_0$
within $0 \leq t \leq t_0$ according to
\begin{equation}
	S_0(t) = \sin\left(\frac{\pi}{2}\frac{t}{t_0}\right).
\end{equation}
To avoid nonlinear backaction of the condensate with the source, we also
introduce a spatial ramping of the interaction strength $g$ according to
\begin{equation}
	g(x)=\frac{1}{2}\left[1+\mathrm{tanh}\left(\kappa(x-x_1)\right)\right] g
\end{equation}
where $\kappa$ and $x_1$ are chosen such that $g(x_0) < 0.01 g$ at the
position $x_0$ of the source and $g(x) > 0.99 g$ within the spatial region
in which the scattering potential $V(x)$ is finite.
We verified that the presence of this ramping, which in practice could be
induced by a spatially varying confinement frequency $\omega_\perp \equiv
\omega_\perp(x)$, does not cause a significant additional amount of
reflection.

To avoid artificial backreflection of atoms from the boundaries of the
numerical grid, we furthermore introduce complex absorbing potentials
of the form $V_{abs}(x) = - i \gamma(x)$ where $\gamma(x)$ is vanishing
throughout the scattering region and adiabatically ramped to a finite
positive value in the vicinity of the boundaries \cite{1986Kosloff,MoiO04JPB}.
As the single-particle Hamiltonian of the system thereby becomes
non-hermitean, care is required for the inclusion of this imaginary potential
into the equation of the one-body density matrix $\Gamma(x_1,x_2)$; we obtain
$-i[\gamma(x)+\gamma(y)]\Gamma(x,y,t)$ as an additional term in
Eq.~(\ref{eq:gamma}).

\section{Transport through double barrier potentials}

\label{sec:db}

Let us first investigate resonant transport of a Bose-Einstein condensate
through a symmetric double barrier potential, which can be interpreted as a
Fabry-Perot resonator for coherent matter waves \cite{CarLar00PRL,Car01PRA}.
Such an atomic quantum dot may exhibit long-lived quasi-bound eigenstates
that allow for perfect transmission at energies in the immediate vicinity
of the eigenenergies of such states, while off-resonant components of the
incident wave will undergo reflection.
In the context of guided atom lasers, such double barrier potentials may
therefore serve as tools for energetic purification of the incident beam,
in the case that the source is, in practice, not yielding a perfectly 
monochromatic matter wave.

In accordance with our previous studies \cite{PauRicSch05PRL}, we consider a
double barrier potential of the form
\begin{equation}
  V(x)=V_0 \left[ e^{-(x-x_1-L/2)^2/\sigma^2} + e^{-(x-x_1+L/2)^2/\sigma^2}
  \right]
  \label{eq:Vdb}
\end{equation}
with $x_1$ the central position between the two barriers, $L$ the
distance between the maxima, and $\sigma$ the width of the barriers.
The Bose-Einstein condensate is assumed to consist of $^{87}\mathrm{Rb}$ atoms
with the scattering length $a_{sc}=5.77$nm and to propagate in a waveguide
with the transverse confinement frequency 
$\omega_\perp=2\pi \times 10^3 s^{-1}$.
In terms of $\omega_\perp$ and the transverse oscillator length 
$a_\perp = \sqrt{\hbar/m\omega_\perp} \simeq 0.34\mu$m, we therefore have
$g \simeq 0.034 \hbar \omega_{\perp} a_\perp$.
As in Ref.~\cite{PauRicSch05PRL}, we choose $V_0=\hbar \omega_\perp$ and 
$L = 10 \sigma = 5 \mu$m $ \simeq 14.7 a_\perp$ for the parameters of the
potential (\ref{eq:Vdb}).

\begin{figure}
\includegraphics[width=1.0\columnwidth]{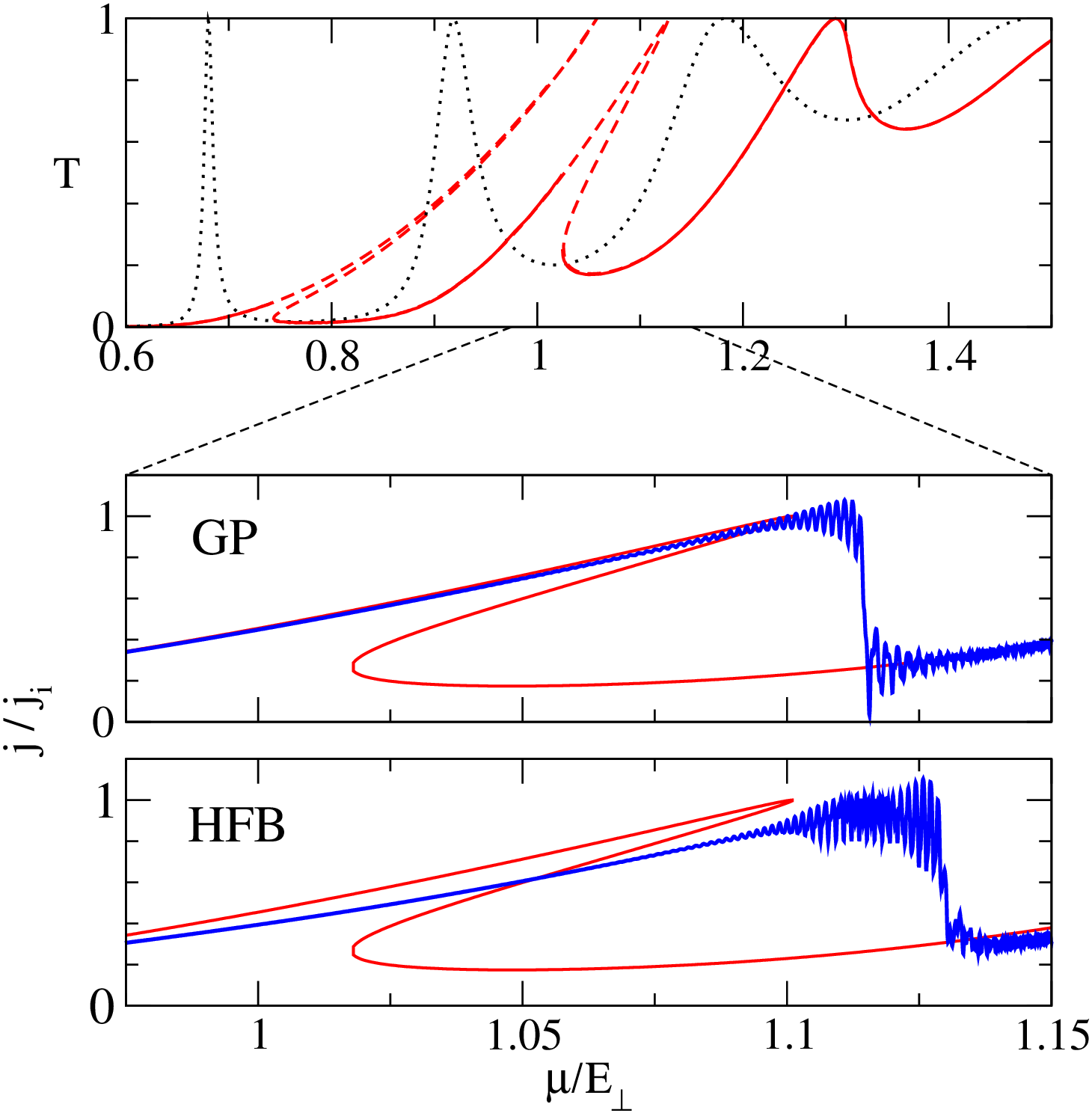}
\caption{\label{fig:transportDB} (Color online)
Time-dependent sweep across a resonance of the double barrier potential 
(\ref{eq:Vdb}) [which is displayed in the lower panel of
Fig.~(\ref{fig:transportDB-density})].
The upper panel shows the nonlinear transmission spectrum at the incident
current $j_i = 1.6\omega_\perp$ obtained from the integration of the
Gross-Pitaevskii equation (\ref{eq:gps}) with the interaction strength 
$g \simeq 0.034 \hbar \omega_{\perp} a_\perp$.
Dashed lines mark the resonance branches that are not directly
accessible by means of a straightforward injection of matter waves at fixed
chemical potential.
The dotted line represents the linear transmission spectrum at $g=0$.
The middle and lower panels show the time-dependent transmission that is
obtained by sweeping the chemical potential according to Eq.~(\ref{eq:mut})
during the propagation process.
This sweep allows one to populate the upper branch of the resonance peak
that corresponds to the 5th excited quasibound state within the cavity.
Apart from a slight shift of the distorted resonance peak (due to a
renormalization of the effective interaction strength), the calculation based
on the dynamical Hartree-Fock-Bogoliubov (HFB) equations
(\ref{eq:psi}-\ref{eq:gamma})  (lower panel) reproduces quite well the
Gross-Pitaevskii (GP) calculation (middle panel).}
\end{figure}

The main characteristics of this transport process can be understood on the
level of the inhomogeneous Gross-Pitaevskii equation
\begin{eqnarray}
  i\hbar\frac{\partial}{\partial t}\psi(x,t) & = &
  H_0^{(x)}\psi(x,t)+g\left|\psi(x,t)\right|^{2}\psi(x,t) \nonumber \\
  & & + S(x,t) \exp(-i \mu t / \hbar) \label{eq:gps}
\end{eqnarray}
which corresponds to the lowest-order truncation of the system of equations
(\ref{eq:psi}--\ref{eq:gamma}).
The upper panel of Fig.~\ref{fig:transportDB} displays the transmission
spectrum through this double barrier potential as a function of the
condensate's chemical potential $\mu$, in the absence and presence of the
interaction between the atoms (dotted and solid/dashed lines, respectively),
which is computed by means of stationary scattering solutions of the
Gross-Pitaevskii equation that exhibit the incident current $j_i = 1.6
\omega_\perp$.
For a noninteracting condensate, a sequence of Breit-Wigner peaks would be
obtained in the transmission, whose positions and widths correspond to the
energies and decay rates, respectively, of the quasi-bound states within the
resonator.
These transmission peaks exist also in the presence of interaction, but become
strongly distorted towards higher chemical potentials (for repulsive
interaction) due to the nonlinearity in the Gross-Pitaevskii equation
\cite{PauRicSch05PRL,PauO07PRA}.
A multivalued transmission spectrum with several bistable branches is thereby  
obtained, which is typical for nonlinear transmission problems and arises also
in nonlinear optics \cite{Boy} as well as in the electronic transport through
quantum wells \cite{GolTsuCun87PRL,PreIonCap91PRB,Azb99PRB}.

It is easy to see that a direct, straightforward injection of matter waves
onto this atomic Fabry-Perot resonator at a given chemical potential $\mu$
would always lead to the population of the lowest branch in the transmission
spectrum \cite{PauRicSch05PRL,PauO07PRA}.
Hence, resonant transmission through long-lived quasi-bound states with narrow
widths cannot be achieved in this way, and the atomic quantum dot
qualitatively acts like a simple potential barrier, giving rise to nearly
perfect (classical-like) reflection of the condensate for chemical potentials
well below the barrier height $V_0$.
Indeed, if the chemical potential of the incident beam is chosen to match the
level of an internal (noninteracting) quasi-bound state, a finite population
of this state will shift the associated level towards higher
energies, and the matter-wave beam will no longer be on resonance.
On the other hand, injecting the condensate at the appropriate chemical
potential that corresponds to the position of the \emph{distorted} resonance
peak would not yield any population in the internal quasi-bound state in the
first place if the resonator was initially empty.

In Ref.~\cite{PauRicSch05PRL} we proposed a control scheme that allows one to 
overcome this limitation and to achieve resonant transport of the condensate
in the presence of finite interaction.
The basic idea is that the chemical potential of the condensate (or,
equivalently, the offset potential of the waveguide) needs to be adiabatically
varied \emph{during the propagation process}, in order to follow the upper
branch of one of the distorted resonance peaks.
The feasibility of this scheme was demonstrated through numerical simulations
on the basis of the time-dependent Gross-Pitaevskii equation.
Starting at a chemical potential on the left shoulder of the resonance peak
corresponding to the 5th excited quasibound state (which is magnified in the
middle and lower panels of Fig.~\ref{fig:transportDB}), and increasing the
chemical potential up to the value that equals the peak's top position, we
obtained nearly perfect transmission on finite time scales, limited only by
the dynamical instability of the nonlinear scattering state at resonance
\cite{PauRicSch05PRL}.

The middle and lower panels of Fig.~\ref{fig:transportDB} display the outcome
of a similar time-dependent transport process, which was calculated here
with the Gross-Pitaevskii equation (middle panel) as well as through the
the numerical integration of the dynamical Hartree-Fock Bogoliubov equations
(\ref{eq:psi}--\ref{eq:gamma}).
In these calculations, we first populated a scattering state at the chemical
potential $\mu_0 = 0.97 \hbar \omega_\perp$, by ramping up the source
amplitude within the propagation time $t_0 = 10^3 \omega_\perp^{-1}$, and then
subjected it to a linear sweep of the chemical potential according to 
\begin{equation}
	\mu(t) = \mu_0 + (t - t_0) \dot{\mu} \label{eq:mut}
\end{equation}
with $\dot{\mu} = 5 \times 10^{-5} \hbar \omega_\perp^2$.
Both calculations lead to a similar scenario, namely the adiabatic population
of the upper branch of the resonance peak corresponding to the 5th excited
quasi-bound state in the resonator, and the subsequent decay to a
low-transmission state at propagation times $t$ at which the corresponding
chemical potential $\mu(t)$ exceeds the position of the resonance peak.
A slight displacement in the peak position and in the slope of the resonance
branch is observed in the HFB calculations when being compared with the
nonlinear transmission curve obtained from the Gross-Pitaevskii equation.
We attribute this to a renormalization of the effective nonlinearity strength
in Eqs.~(\ref{eq:psi}--\ref{eq:gamma}) due to the coupling of the condensate
wavefunction to the two-point functions $\Phi(x_1,x_2,t)$ and 
$\Gamma(x_1,x_2,t)$.

\begin{figure}
\includegraphics[width=1.0\columnwidth]{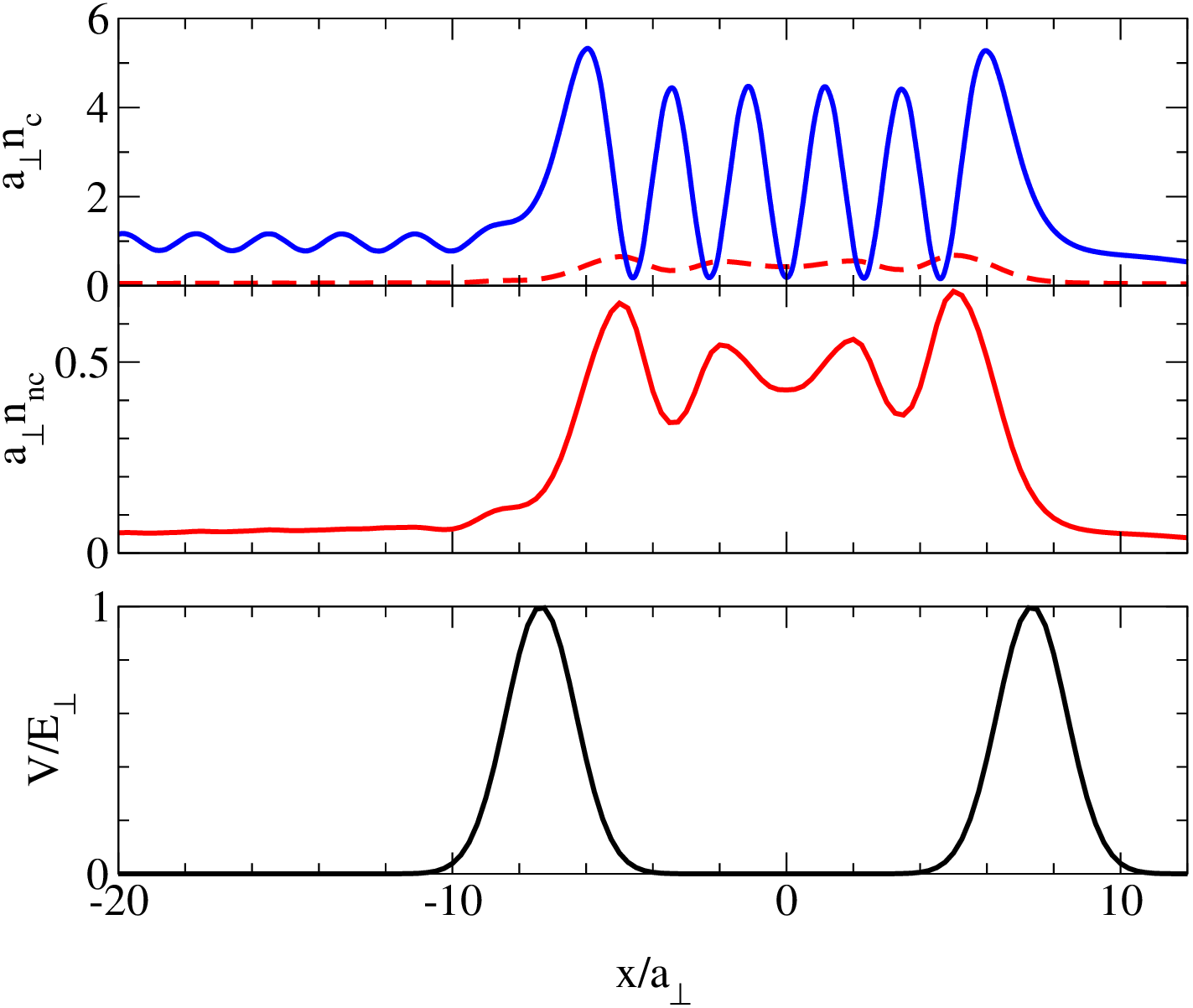}
\caption{\label{fig:transportDB-density} (Color online)
Density of the condensate wavefunction (upper panel) and of the noncondensed
atoms (middle panel, also shown as red dashed line in the upper panel)
near a resonance of the double barrier potential (\ref{eq:Vdb}).
The plot represents a snapshot of the time-dependent sweep across the
resonance shown in Fig.~\ref{fig:transportDB} at the evolution time at which
the chemical potential equals $\mu = 1.1$, corresponding to a transmission of 
$T \simeq 0.85$.
The lower panel shows the double barrier potential.
}
\end{figure}

Fig.~\ref{fig:transportDB-density} displays the condensate density ($n_c$) as
well as the density of noncondensed atoms ($n_{nc}$) of the instantaneous
scattering state at the propagation time $t$ with $\mu(t) = 1.1$ at which an
overall transmission of $T \simeq 0.85$ is obtained.
We clearly see a finite cloud of noncondensed atoms within the resonator,
which is obviously induced by the large local density fluctuations of the
condensate.
The noncondensed denisty is found to slowly increase with time during the sweep
of the chemical potential, but remains rather low compared to the condensate
density on the time scales that are considered here.

Our study indicates that the control scheme proposed in
Ref.~\cite{PauRicSch05PRL} is working not only within the Gross-Pitaevskii
description of the transport process, but remains feasible also from a more
microscopic point of view taking into account the noncondensed cloud that is
generated within the scattering region.
We remark, however, that this conclusion strongly depends on the specific
parameters at hand.
Indeed, if we increase the microscopic interaction strength of the atoms by a
factor two (which could possibly be achieved through a Feshbach resonance)
while keeping the other parameters of the system fixed, we find rather strong
depletion of the condensate within the resonator during the time-dependent
sweep of the chemical potential across the resonance.
In this situation, the truncation scheme leading to the equations
(\ref{eq:psi}--\ref{eq:gamma}) would not be justified any longer, and we expect 
that at that particular high value of the interaction strength the overall
coherence of the atomic cloud cannot be maintained and the condensate would
become destroyed during the transport process.

\section{Transport through disorder potentials}
\label{sec:dis}

Let us now consider the transport of Bose-Einstein condensates through
disorder potentials of finite range, consisting of a more or less random
sequence of several barriers and wells with different heights and depths.
Such disorder potentials received special attention in the past years due to
the possibility of realizing Anderson localization with ultracold atoms
\cite{CleO05PRL,ForO05PRL,SchO05PRL,SanO07PRL,SkiO08PRL,BilO08N,RoaO08N}.
This task was ultimately achieved in two independent experiments
\cite{BilO08N,RoaO08N} studying the expansion of condensates within optical
speckle fields \cite{BilO08N} as well as within bichromatic optical lattices
\cite{RoaO08N}, in a parameter regime where effects of the atom-atom
interaction on the expansion process could be neglected.
Indeed, various theoretical studies revealed that the presence of a finite
nonlinearity in the wave equation that describes the expansion process, which
would correspond to a finite interaction strength within the mean-field
description of the condensate, would lead to a significant deviation from the
scenario of Anderson localization, involving indefinite spreading of the
tails of the condensate wavefunction instead of exponential localization
\cite{She93PRL,KotWei04PRL,PikShe08PRL,KopO08PRL}.

The effect of a finite interaction is predicted to be even more dramatic in
\emph{scattering} processes through disorder potentials.
Here, Anderson localization would manifest in an exponential descrease of the
average transmission $T$ with the spatial length $L$ of the disorder region
(or, more precisely, in a linear increase of the average $\langle -\mathrm{ln}
T \rangle$ with $L$), which is induced by the destructive interference between
multiply reflected components of the scattering wavefunction \cite{AndO80PRB}.
In the presence of a finite nonlinearity, however, a crossover to an
\emph{algebraic} decrease of the average transmission according to $\langle T
\rangle \sim (L + L_0)^{-1}$ is encountered \cite{PauO05PRA}, which would
essentially correspond to Ohm's law for incoherent transport.
We find that this crossover is directly connected with the appearance of
\emph{permanently time-dependent scattering} of the condensate \cite{PauO05PRA,PauO07PRL}.
That is, the injection process of the condensate according to the working
principle of the guided atom laser does not lead to the population of a
quasi-stationary scattering state, but induces time-dependent fluctuations of
the condensate density beyond a certain critical value of the source amplitude.

\begin{figure}
\includegraphics[width=1.0\columnwidth]{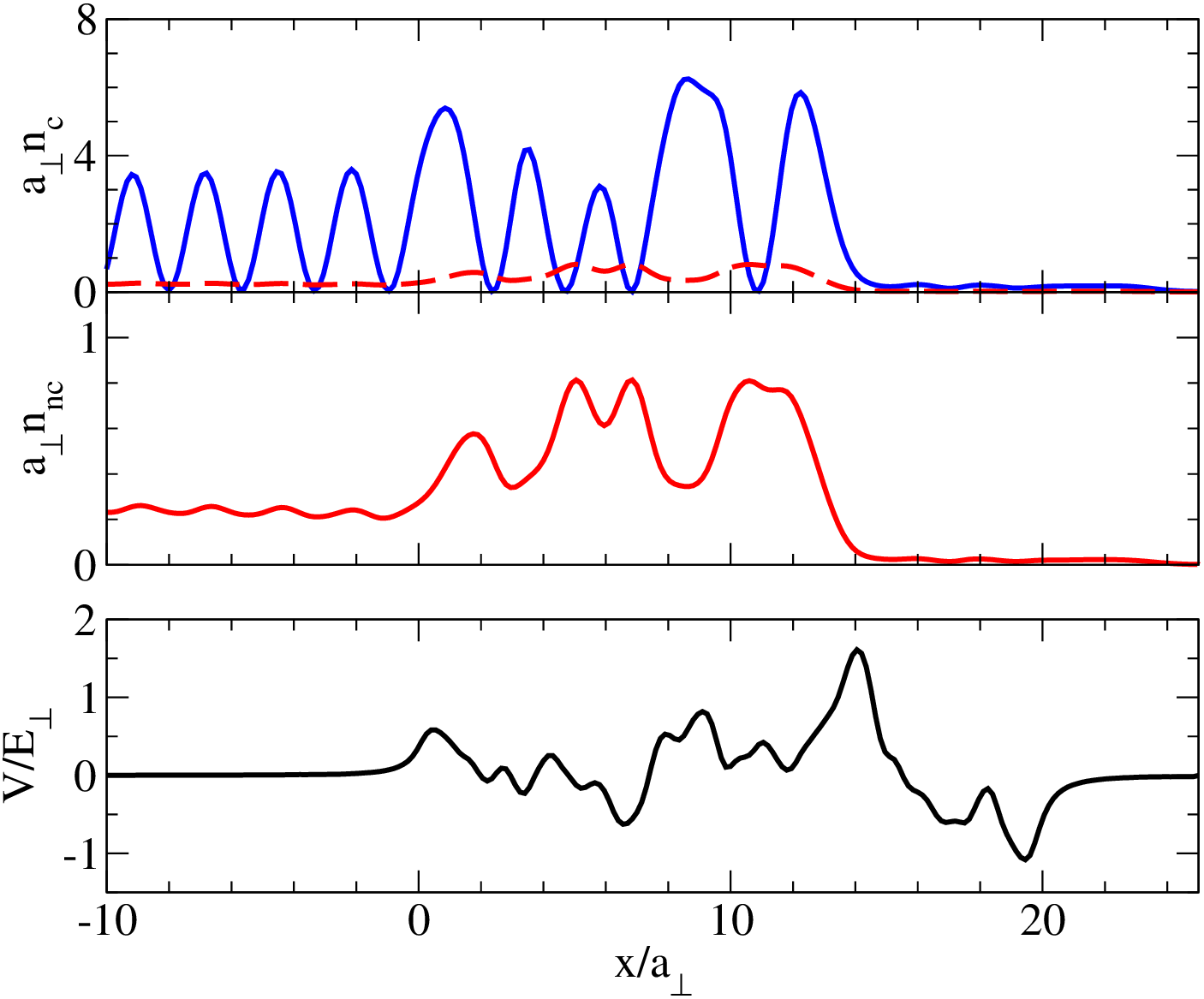}
\caption{\label{fig:disorder-density} (Color online)
Density of the condensate wavefunction (upper panel) and of the noncondensed
atoms (middle panel, also shown as red dashed line in the upper panel)
during the time-dependent propagation process of the condensate through a
disorder potential, at the evolution time $t = 1000 /\omega_\perp$
(corresponding to $160$ ms).
The incident current $j_i = 1.5 \omega_\perp$ was chosen such that a
Gross-Pitaevskii calculation of the transport process yields a
quasi-stationary scattering state.
During the propagation process, noncondensed atoms are found to slowly
accumulate preferrably around the minima of the disorder potential, which is
shown in the lower panel.
The total depletion, however, is still relatively low at 
$t = 1000 / \omega_\perp$.
}
\end{figure}

\begin{figure}
\includegraphics[width=1.0\columnwidth]{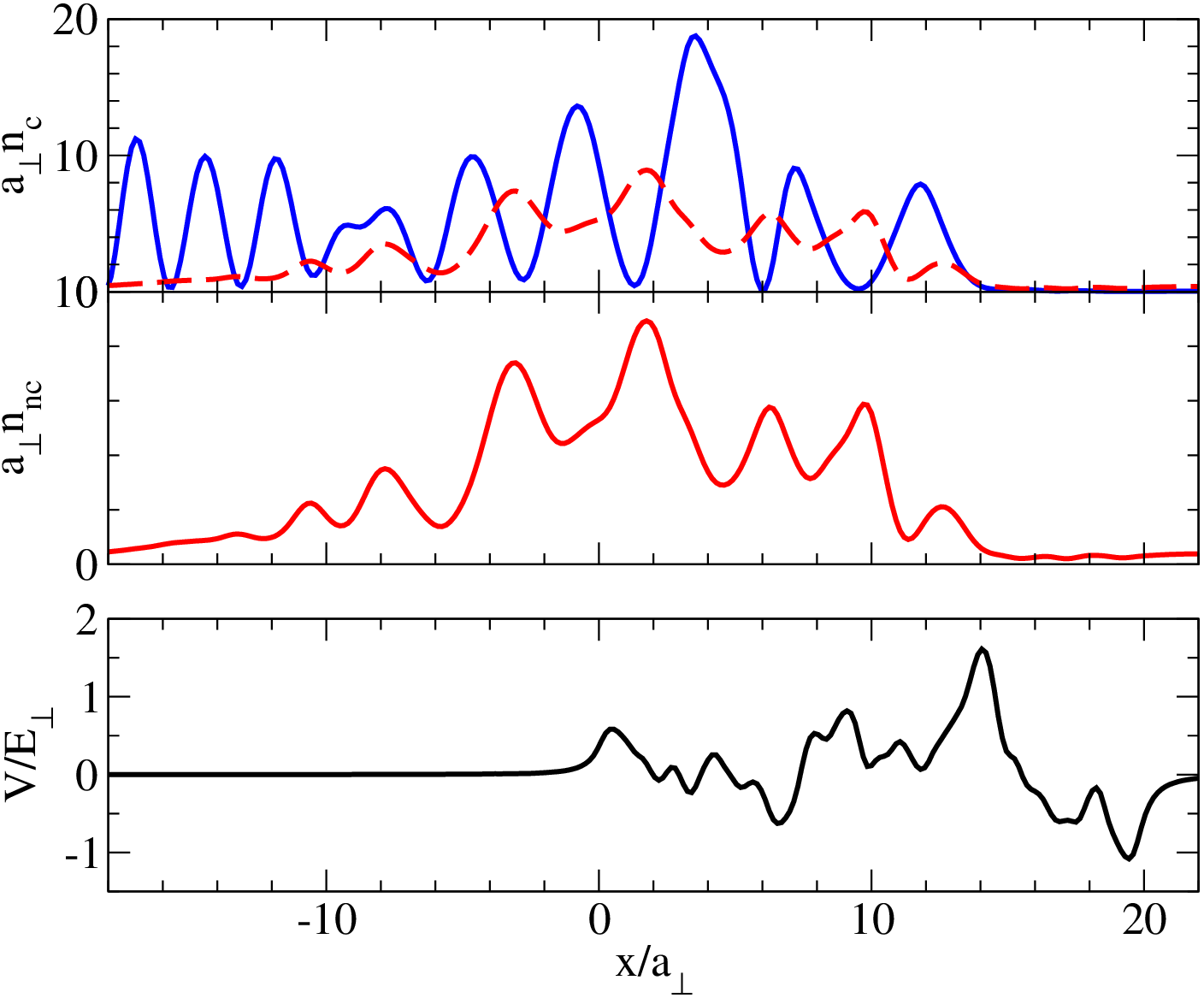}
\caption{\label{fig:fluctuating-density} (Color online)
Same as Fig.~\ref{fig:disorder-density} for a higher incident current
$j_i = 4.775 \omega_\perp$ for which the Gross-Pitaevskii calculation yields
permanently time-dependent scattering.
The figure shows a snapshot at the evolution time $t = 500 /\omega_\perp$.
Here the density of noncondensed atoms $n_{nc}$ is of the same order as the 
condensate density, which indicates strong depletion and the destruction of
the condensate on the microscopic level.
}
\end{figure}

The main finding of our investigation of this problem with the kinetic
Hartree-Fock-Bogoliubov equations (\ref{eq:psi}--\ref{eq:gamma}) is that the
appearance of permanently time-dependent scattering within the Gross-Pitaevskii
description of the transport process involves strong depletion and the
production of a comparatively large cloud of noncondensed atoms.
This indicates that the overall coherence characterizing the condensate
becomes destroyed on a microscopic level, which would indeed be consistent
with an Ohm-like scenario of incoherent transport through the disorder region.
We exemplify this behaviour with one particular, relatively short disorder
potential, displayed in the lower panels of Figs.~\ref{fig:disorder-density}
and \ref{fig:fluctuating-density}.
For the chemical potential of the incident beam, we consider $\mu = 1.0 \hbar
\omega_\perp$ which is of the same order as the barrier heights of the
disorder potential.
As in Section \ref{sec:db}, we choose $\omega_\perp=2\pi \times 10^3 s^{-1}$
for the transverse confinement frequency of the waveguide and 
$g = 0.034 \hbar \omega_{\perp} a_\perp$ for the effective
one-dimensional interaction strength.

Let us first consider the transport process across this disorder potential at
the incident current $j_i = 1.5 \omega_\perp$ for which a Gross-Pitaevskii
calculation yields quasi-stationary scattering.
At this relatively low value of the incident current, we find nearly stationary
dynamics on the level of the kinetic Hartree-Fock-Bogoliubov equations
(\ref{eq:psi}--\ref{eq:gamma}) as well, where the density of the scattering
state is, as shown in Fig.~\ref{fig:disorder-density}, dominated by the
coherent component associated with the condensate.
As for the double barrier potential discussed in Section \ref{sec:db}, a
finite, slowly increasing amount of depletion is generated within the
scattering region, with noncondensed atoms preferrably accumulating around the
combined minima of the disorder potential and the condensate density.
On time scales of the order of $1000 \omega_\perp^{-1}$ corresponding to
$160$ ms, however, the total density of noncondensed atoms is still rather low
compared to the condensate, which indicates that the coherence of the atomic
beam should be well preserved also on the microscopic level.

The situation is rather different for a high incident current such as 
$j_i = 4.775 \omega_\perp$, for which the simulation of the scattering process
on the basis of the Gross-Pitaevskii equation yields permanently time-dependent
fluctuations.
Such time-dependent behaviour is also found on the level of the kinetic
Hartree-Fock-Bogoliubov equations (\ref{eq:psi}--\ref{eq:gamma}), together
with a rather large amount of depletion, as shown by the snapshot displayed in
Fig.~\ref{fig:fluctuating-density}.
In that case, the truncation scheme leading to the closed system of equations
(\ref{eq:psi}--\ref{eq:gamma}) can no longer be justified, and higher-order
cumulants, such as $\langle \delta \hat{\psi}(x_1,t) \delta \hat{\psi}(x_2,t)
\delta \hat{\psi}(x_3,t) \rangle$ for instance, would have to be taken into
account as well for an accurate description of the transport process.
We conjecture that in this situation the condensate fraction becomes destroyed
on a microscopic level, and the transport process becomes incoherent.

\begin{figure}
\includegraphics[width=1.0\columnwidth]{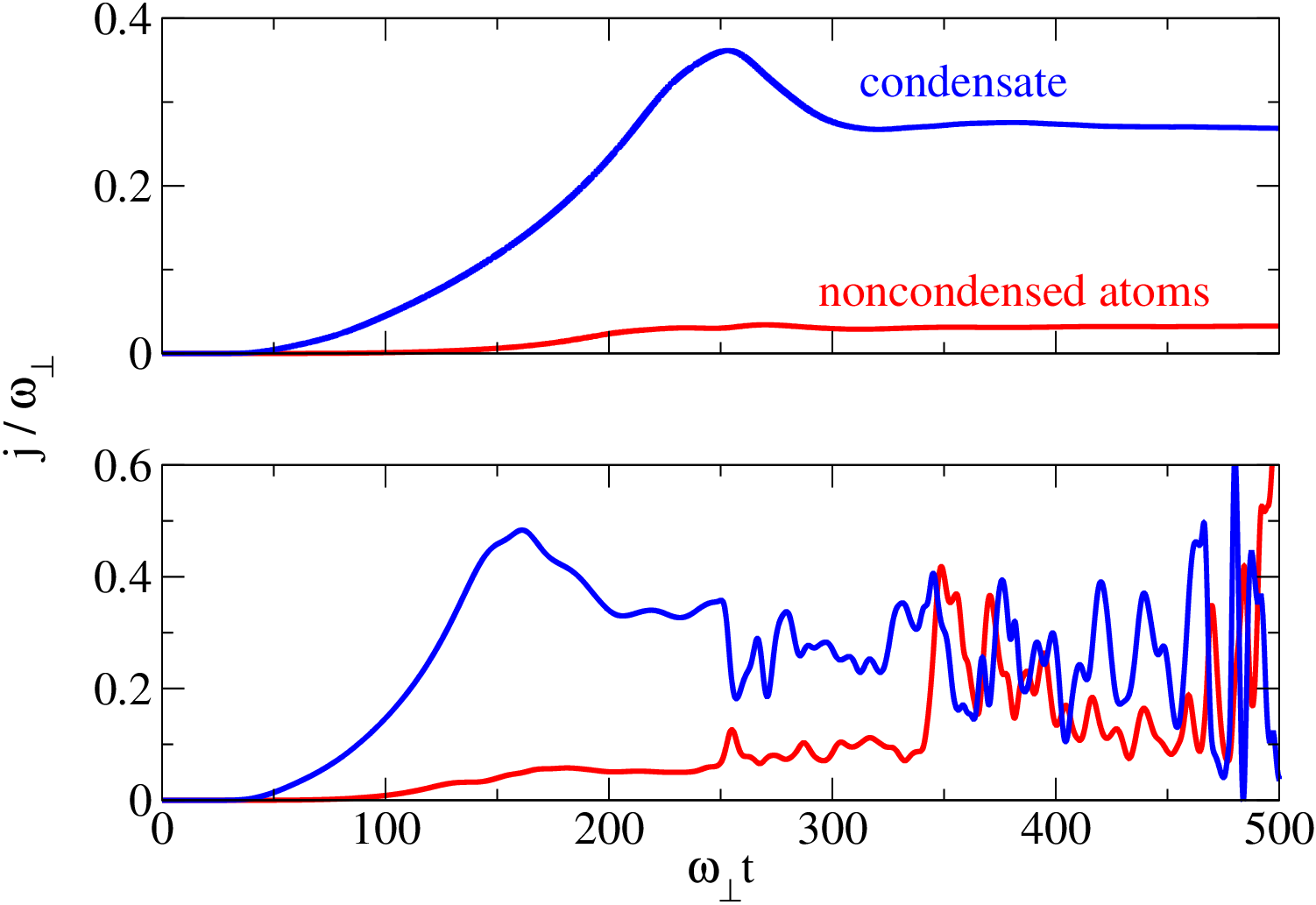}
\caption{\label{fig:current} (Color online)
Total current of the condensate and the noncondensed atoms (blue and red
lines, respectively) as a function of the propagation time, for the incident
currents $j_i = 1.5 \omega_\perp$ (upper panel) and $j_i = 4.775 \omega_\perp$
(lower panel), in the presence of the disorder potential that is shown in
Fig.~\ref{fig:disorder-density}.
The plots include the time interval of the initial ramping process of the
source amplitude, which lasts until $t \simeq 300 / \omega_\perp$.
In the case of a rather low incident current (upper panel) for which a
Gross-Pitaevskii calculation would yield quasi-stationary transport, 
depletion remains rather low on finite time scales.
At higher incident currents for which permanently time-dependent scattering
manifests on the Gross-Pitaevskii level, we find strongly enhanced depletion
and a current of noncondensed atoms that is comparable to the current of the
condensate fraction.
This indicates, on a microscopic level, that the condensate becomes destroyed
during this time-dependent scattering process.
}
\end{figure}

Fig.~\ref{fig:current} provides a comparison of the time-dependent currents of
the condensate and the noncondensed atoms [cf.\ Eq.~(\ref{eq:current})] for
the cases of low and high intensity of the incident matter-wave beam.
While a stable stationary flow, dominated by the coherent component of the
condensate, is encountered in the presence of a low incident current,
time-dependent dynamics arises already during the ramping process of the
source for $j_i = 4.775 \omega_\perp$ and induces in turn rather strong
depletion from $\omega_\perp t \simeq 350$ on.
Obviously, our calculation is, in this latter case, not trustable any longer
beyond that evolution time as far as the accurate prediction of the
time-dependent densities and currents of the condensate and the noncondensed
fractions is concerned.

\begin{figure}
\includegraphics[width=1.0\columnwidth]{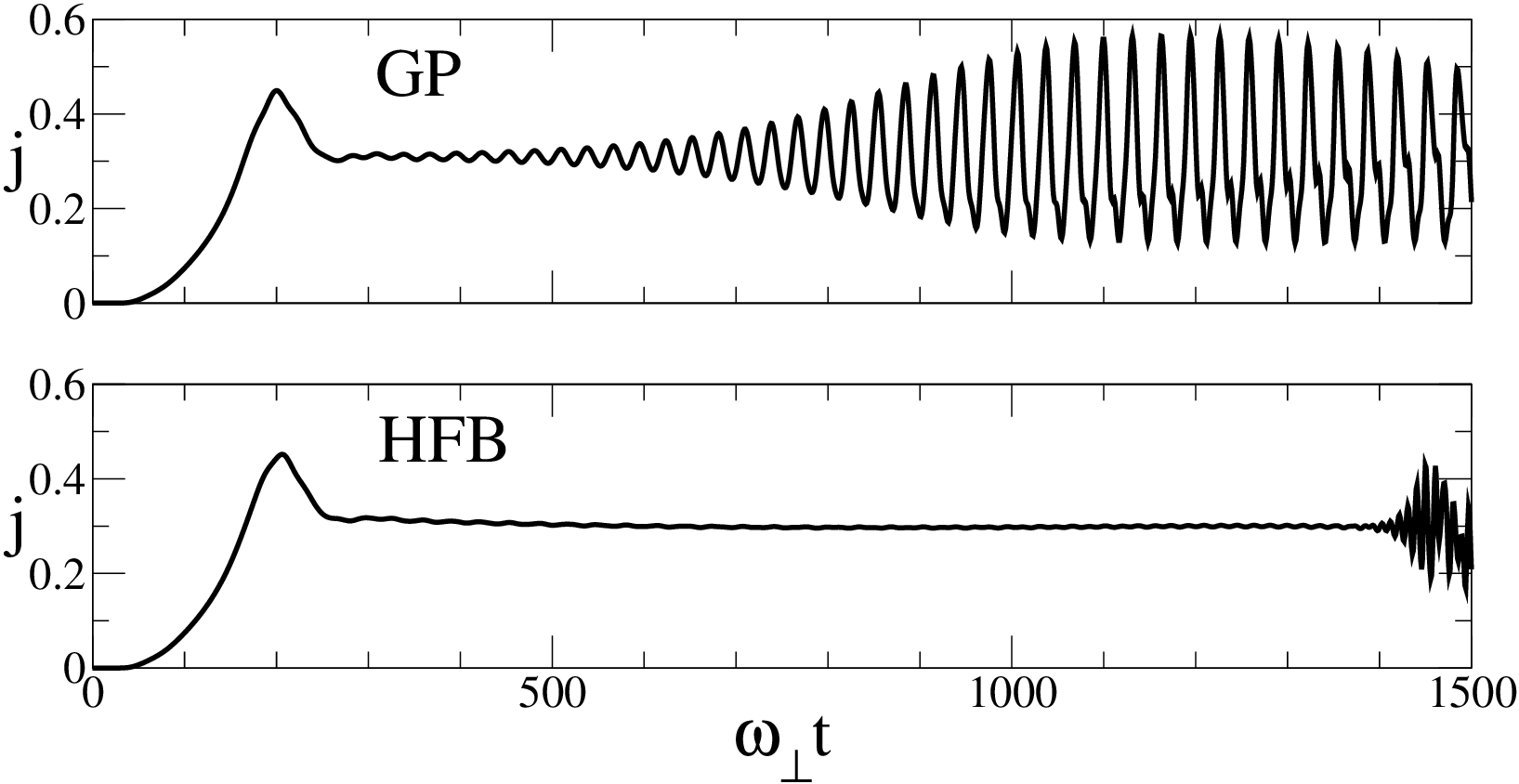}
\caption{\label{fig:disorder-onset}
Total atomic current across the disorder potential as a function of time,
calculated for the critical incident current $j_i=2.387 \omega_\perp$ at which
time-dependent scattering begins to appear on the level of the Gross-Pitaevkii
equation (upper panel).
A simulation of this scattering process with the kinetic
Hartree-Fock-Bogoliubov equations (lower panel), however, yields
quasi-stationary behaviour on a time scale of the order of $1000
\omega_\perp^{-1}$, with a reasonably low amount of depletion
(between 15 and 20 percent of the total density), while dynamical instability
and strong depletion sets in at $\omega_\perp t \simeq 1500$.}
\end{figure}

The scenario discussed here is systematically found also for a number of other
disorder potentials of similar length as the one shown in
Figs.~\ref{fig:disorder-density} and \ref{fig:fluctuating-density}.
Stationary scattering on the Gross-Pitaevskii level implies, in general, 
a nearly stationary and fairly coherent flux on the level of the kinetic
Hartree-Fock-Bogoliubov equations, whereas time-dependent scattering within
the Gross-Pitaevskii equation leads to strong depletion.
We point out, however, that there is no rigorous one-to-one relation between
these two phenomena. 
Indeed, exceptions from this general rule particularly arise if the incident
current of the condensate is close to the critical value $j_i^{(c)}$ beyond
which permanently time-dependent scattering occurs on the level of the
Gross-Pitaevskii equation.
In this case, a small modification of the effective nonlinearity, caused by
the coupling to the noncondensed cloud, can have a drastic effect and may
destabilize --- or, in some cases, also re-stabilize --- the flow of the
condensate.
An example is shown in Fig.~\ref{fig:disorder-onset}, where the time
evolution of the current across the disorder potential under consideration is
shown for the critical incident current $j_i=2.387 \omega_\perp$
beyond which the flux becomes unstable on the Gross-Pitaevskii level.
In contrast to the Gross-Pitaevskii calculation, the simulation of this
transport process with the kinetic Hartree-Fock-Bogoliubov equations does not
give rise to time-dependent fluctuations of the current on time scales of the
order of $1000 \omega_\perp^{-1}$ (which would correspond to $160$ ms) and
leads to a stable flux with a reasonably low amount of depletion, until
destabilization sets in later on at $\omega_\perp t \simeq 1500$.

\section{\label{sec:concl}Conclusion}

In this work, we investigated the transport of guided bosonic matter waves
through one-dimensional scattering potentials by means of the inhomogeneous
Hartree-Fock-Bogoliubov equations, which simulate the coherent injection
of those matter waves from a trapped Bose-Einstein condensate
\cite{GueO06PRL}.
A straightforward numerical integration of these Hartree-Fock-Bogoliubov
equations yields nearly stationary, mean-field-like scattering dynamics
of the bosonic flow on reasonably long time scales, provided the effective
one-dimensional interaction strength is not too large and the integration of
the corresponding Gross-Pitaevskii equation leads to quasistationary
scattering.
In that case, a comparatively low amount of depletion is obtained, with
noncondensed atoms preferrably accumulating around the local minima of the
scattering potential.
If, on the other hand, the nonlinear scattering dynamics of the condensate is
explicitly time-dependent on the level of the Gross-Pitaevskii equation, we
encounter strong depletion and a rather large density of noncondensed atoms
within the Hartree-Fock-Bogoliubov description.
This indicates that the condensate becomes destroyed on a microscopic
many-body description of the transport process, and that the atoms are
individually propagating across the scattering potential.

It would be interesting, though beyond the scope of this work, to compare our
findings with the truncated Wigner approach
\cite{GarZol,SteO98PRA,SinLobCas01PRL}, which represents an alternative method
to describe dynamical transport processes of Bose-Einstein condensates beyond
the Gross-Pitaevskii description.
In this approach, the microscopic quantum dynamics of the atomic cloud is
described by a set of classical fields which sample the distribution of the
initial many-body state describing the Bose gas under consideration, and which
evolve according to the time-dependent Gross-Pitaevskii equation.
In our situation, this initial state would correspond to an almost perfect
Bose-Einstein condensate of the ``reservoir'' atoms that are confined in the
three-dimensional magnetic trap.
The classical fields that would have to be used for the truncated Wigner
approach would therefore be rather similar at the initial stage of the
matter-wave injection process, and deviations between them, which ultimately
give rise to a loss of coherence, should mainly arise in the presence of
strongly time-dependent scattering where the occurrence of classical chaos in
the Gross-Pitaevskii equation might induce an exponential sensitivity of the
time evolution with respect to initial conditions.
This would be consistent with our observation that depletion is particularly
strong in the regime of permamently time-dependent scattering.

An interesting study in this context was recently reported in
Ref.~\cite{ScoHut08PRA}, where the quantitative amount of depletion was
computed for a homogeneously moving three-dimensional condensate (initially
defined as a plane wave in a box with periodic boundaries) in the presence of
an obstacle or a disorder potential.
Using the truncated Wigner method, the authors showed that nonstationary,
turbulent flow of the condensate leads to rather rapid depletion, on time
scales that seem to be roughly comparable with our investigations
\cite{timescale}.
Good agreement between the time-dependent Hartree-Fock-Bogoliubov approach and
the truncated Wigner approach was, moreover, reported in other studies
\cite{WueO07PRA} that were focusing on collapsing Bose-Einstein condensates.

We finally point out again that several idealizations and approximations were
employed for the desciption of the bosonic transport process.
Specifically, we assumed a perfect condensate in the reservoir, which is not
significantly depopulated during the outcoupling process, we neglected the
(virtual) population of transversally excited modes in the waveguide, and we
excluded nonlinear back-action with the source assuming that the interaction
strength of the waveguide atoms is negligible in the vicinity of the magnetic
trap.
These simplifications would have to be reconsidered for the simulation of a
realistic matter-wave propagation experiment based on an atom laser.
We do, however, not expect that the general findings obtained in this study
would qualitatively alter if these complications are properly taken into
account.

\begin{acknowledgments}

T.P. gratefully acknowledges funding by the Excellence Initiative by the German
Research foundation (DFG) through the Heidelberg Graduate School of
fundamental Physics (grant number GSC 129/1), the Global Networks Mobility
Measures the Frontier Innovation Fund of the Univeristy of Heidelberg, and
the Alexander von Humboldt Foundation.
P.S. acknowledges support through the DFG Forschergruppe FOR760.
We are, furthermore, indepted to Thomas Gasenzer, David Hutchinson, Gora
Shlyapnikov, and Sebastian W\"uster for enlightening discussions.

\end{acknowledgments}


\begin{thebibliography}{10}
\bibitem{timescale}
In Ref.~\cite{ScoHut08PRA}, the characteristic time scale for the loss of
  coherence was found to be of the order of 1 ms, which is smaller compared to
  our findings. We attribute this to the considerably larger three-dimensional
  atomic densities that were considered in Ref.~\cite{ScoHut08PRA}.
\end{thebibliography}
\end{document}